\documentclass[twocolumn,showpacs]{revtex4}
\usepackage{graphicx}
\usepackage{dcolumn}
\usepackage{bm}

\begin{document}

\title{Fluid Mechanical and Electrical\\
Fluctuation Forces in Colloids}

\author{D. Drosdoff}
\author{A. Widom}
\affiliation{Physics Department, Northeastern University, Boston MA 02115}

\begin{abstract}
Fluctuations in fluid velocity and fluctuations in electric fields
may both give rise to forces acting on small particles in colloidal
suspensions. Such forces in part determine the thermodynamic
stability of the colloid. At the classical statistical thermodynamic
level, the fluid velocity and electric field contributions to
the forces are comparable in magnitude. When quantum fluctuation effects
are taken into account, the electric fluctuation induced
van der Waals forces dominate those induced by purely fluid mechanical
motions. The physical principles are applied in detail for the case
of colloidal particle attraction to the walls of the suspension container
and more briefly for the case of forces between colloidal particles.
\end{abstract}

\maketitle

\section{Introduction \label{Intro}}

Fluctuations in thermodynamic field parameters may give rise to
forces acting on colloidal particles. Examples of such field parameters
include velocity fields within the fluid surrounding the particle and local
electric fields. Our purpose is to examine the magnitudes of these
two force effects. We seek to ascertain which of these forces have
the dominant effect on the thermodynamic properties of colloidal
suspensions.

The classical statistical thermodynamic contribution to fluctuations forces
are scaled by the thermal energy \begin{math} k_BT \end{math}.
The forces due to quantum mechanical zero point fluctuations
involve a frequency scale \begin{math} \omega_\infty \end{math} of
motion and are thereby scaled by the energy quantum
\begin{math} \hbar \omega_\infty \end{math}.
The total magnitude of such forces depend
upon whether or not the fluctuations are
classical or quantum mechanical in nature. We find for temperatures in
the neighborhood of room temperature that the electric field
fluctuation force\cite{Casimir:1948} dominates the hydrodynamic fluctuation
force\cite{Ivlev:2002} in the fully quantum mechanical theory.

The Einstein theory of purely {\em classical} statistical thermodynamic
fluctuation forces will be reviewed in Sec.\ref{StatThermo}. The general
results are employed for the specific example of the force on a colloidal
spherical particle due to a neighboring hard wall. The hydrodynamic case is
discussed in Sec.\ref{fluid} while the electric field case is discussed
in Sec.\ref{DipMom}. The general theory of quantum mechanical fluctuations
are explored in Sec.\ref{quant} and the frequency scales of both
fluid mechanical and electrical fluctuations are considered in
Sec.\ref{frequency}. The frequency scales are such that the fluid mechanical
fluctuation forces are classical while the electric dipole fluctuation
forces are quantum mechanical. The latter thereby dominate the former as
discussed in Sec.\ref{QDF}. Although the electric field {\em static} quantum
fluctuation forces dominate the classical fluid {\em static} fluctuation
forces, the fluid forces are nonetheless observable. If the bandwidth of
experimental observations of Brownian motion coincides
with a frequency regime wherein fluid mechanics holds true, then fluid
fluctuation forces may be (and have been) measured. This point is discussed in
Sec.{\ref{fm}} wherein the formula for the fluid and electrical fluctuation
forces are exhibited for two well separated spheres.
In the concluding Sec.{\ref{conc}}, the electric fluctuation forces are shown
to more strongly determine the phase properties of a colloid, i.e. whether
the colloidal particles will form a smooth colloidal
suspension or whether the colloidal particles will undergo phase separation.

\section{Statistical Thermodynamics \label{StatThermo}}

Let us consider two sets of coordinates,
\begin{math} Q=(Q_1,\ldots ,Q_m) \end{math}
and \begin{math} X=(X_1,\ldots ,X_n) \end{math}.
For the moment, let us fix \begin{math} Q \end{math}
and assume that \begin{math} X \end{math} undergoes
classical thermal fluctuations\cite{Landau:1999}.
The Einstein probability\cite{Einstein:1956} \begin{math} P \end{math}
of exhibiting a deviation
\begin{math} \Delta X=(\Delta X_1,\ldots ,\Delta X_n) \end{math}
from thermal equilibrium has a Gaussian form determined by an activation
free energy \begin{math} \Delta F  \end{math}. The activation free energy
is the minimum isothermal work done on the system by the environment in order
to to produce the fluctuation \begin{math} \Delta X \end{math};
Whence it follows that
\begin{eqnarray}
P & \propto & e^{-\Delta F/k_BT},
\nonumber \\
\Delta F & = & \frac{1}{2}
\sum_{j,k}^n G^{-1}_{\ \ jk}(Q)\Delta X_j \Delta X_k ,
\label{StatThermo1}
\end{eqnarray}
wherein the matrix \begin{math} ||G_{jk}(Q) || \end{math} describes
an effective Hook's law compliance. The static classical
fluctuation-response theorem dictates that
\begin{equation}
\overline{\Delta X_i \Delta X_j}=k_BT G_{ij}(Q),
\label{StatThermo2}
\end{equation}
which follows directly from Eq.(\ref{StatThermo1}). The forces
\begin{math} f=(f_1,\ldots ,f_m) \end{math}
conjugate to \begin{math} Q=(Q_1,\ldots ,Q_m) \end{math} may be
derived from the free energy
\begin{equation}
f_l=-\frac{\partial \Delta F}{\partial Q_l}=-\frac{1}{2}\sum_{j,k}^n
\frac{\partial G^{-1}_{\ \ jk}(Q)}{\partial Q_l}\Delta X_j \Delta X_k
\label{StatThermo3}
\end{equation}
The mean value of this fluctuation force
\begin{equation}
\bar{f}_l=-\frac{1}{2}
\sum_{j,k}^n\frac{\partial G^{-1}_{\ \ jk}(Q)}{\partial Q_l}
\overline{\Delta X_j \Delta X_k}
\label{StatThermo4}
\end{equation}
may be evaluated via Eq.(\ref{StatThermo2}) as
\begin{eqnarray}
\bar{f}_l&=&-\frac{k_BT}{2}\sum_{j,k}^nG_{kj}(Q)
\frac{\partial G^{-1}_{\ \ jk}(Q)}{\partial Q_l},
\nonumber \\
\bar{f}_l&=&
\frac{k_BT}{2}\frac{\partial}{\partial Q_l}\ln\left(\det ||G_{jk}(Q) ||\right) .
\label{StatThermo5}
\end{eqnarray}
Employing a reference matrix \begin{math} ||G^{(0)}_{\ \ jk} || \end{math}
which is {\em independent} of \begin{math} Q \end{math}, we may write
Eq.(\ref{StatThermo5}) in terms of an effective potential
\begin{math} U(Q) \end{math}, i.e.
\begin{eqnarray}
\bar{f}_l &=& -\frac{\partial U(Q)}{\partial Q_l},
\nonumber \\
U(Q) &=& \frac{k_BT}{2} \ln \det\left(G^{-1}(Q)G^{(0)}\right).
\label{StatThermo6}
\end{eqnarray}
If the \begin{math} G(Q) \end{math} obeys the Green's function
matrix equation
\begin{equation}
G(Q)=G^{(0)}+G^{(0)}\Sigma (Q)G(Q),
\label{StatThermo7}
\end{equation}
then
\begin{eqnarray}
U(Q) &=& \frac{k_BT}{2} \ln \det\left(1-G^{(0)}\Sigma (Q)\right)
\nonumber \\
U(Q) &=&-\frac{k_BT}{2}\sum_{n=1}^\infty
\frac{1}{n}Tr\left(G^{(0)}\Sigma (Q)\right)^n.
\label{StatThermo8}
\end{eqnarray}

\begin{figure}[tp]
\scalebox {0.7}{\includegraphics{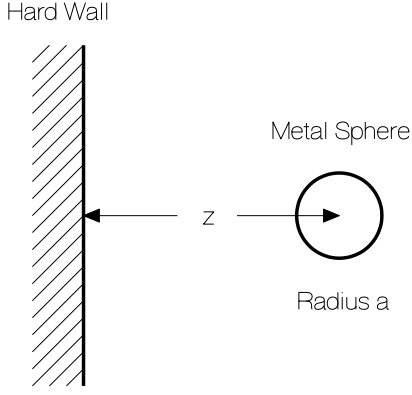}}
\caption{\it Shown is a metal sphere of radius $a$ submerged in a fluid
at a distance $z$ from a hard wall. We seek to compute the fluctuation
forces on the sphere due to the presence of the wall in part
to establish the physically dominant contributions.}
\label{fig1}
\end{figure}

The central result of this section resides in Eqs.(\ref{StatThermo6}),
(\ref{StatThermo7}) and (\ref{StatThermo8}) which describe the effective potential
\begin{math} U(Q) \end{math} of {\em classical} statistical thermodynamic
fluctuation forces in terms of the determinants of the compliance
matrices. The quantum mechanical version of fluctuation forces will
be briefly discussed in the following work where we will explicitly
compute the situation shown in Fig.\ref{fig1}.
We consider two contributions to the fluctuation forces between the sphere
and the wall, namely (i) fluid velocity fluctuations and (ii) electric field
fluctuations.

\section{Fluid Velocity Fluctuations \label{fluid}}

For a single metal sphere (in a fluid) with momentum \begin{math} {\bf p} \end{math}
and position \begin{math} {\bf r} \end{math}, the minimum isothermal work
required to produce the momentum is given by the total kinetic energy
\begin{equation}
\Delta F={\bf p}\cdot \frac{\sf 1}{2{\sf M}(\bf r)}\cdot {\bf p}
=\frac{1}{2}\sum_{j,k}^3 M^{-1}_{\ \ jk}({\bf r})p_j p_k,
\label{fluid1a}
\end{equation}
where the mass matrix \begin{math} ||M_{jk}({\bf r})|| \end{math}
plays the role of the compliance in Eq.(\ref{StatThermo1}).
The static fluctuation response Eq.(\ref{StatThermo2}) now reads
as the equipartition theorem
\begin{equation}
\overline{p_j p_k}=k_BTM_{jk}({\bf r}).
\label{fluid1b}
\end{equation}
Let us consider the mass matrix \begin{math} ||M_{jk}({\bf r})|| \end{math}
in more detail.

If the sphere were very far from the wall \begin{math} z=\infty \end{math},
then the mass matrix would be given by
\begin{equation}
M^{(0)}_{\ \ jk}=\delta_{jk}\left(M+\mu \right)=
\delta_{jk}\left(M+\frac{2\pi \rho a^3}{3} \right),
\label{fluid2}
\end{equation}
wherein \begin{math} \rho  \end{math} is the mass density of the fluid
and \begin{math} M \end{math} is the mass of the sphere.
The Euler mass \begin{math} \mu  \end{math}, which is half the mass of the
displaced fluid, has the following physical interpretation\cite{Landau:2002}.
The kinetic energy of a sphere moving slowly with velocity
\begin{math} {\bf v} \end{math} through an infinite bulk fluid is given by
\begin{eqnarray}
K_{total}&=&K_{particle}+K_{fluid}=
\frac{1}{2}\sum_{j,k}^3 M^{(0)}_{\ \ jk}v_j v_k
\nonumber \\
K_{total}&=&
\frac{1}{2}M|{\bf v}|^2+\frac{1}{2}\mu |{\bf v}|^2.
\label{fluid3}
\end{eqnarray}
The mass \begin{math} M \end{math} enters into the particle
kinetic energy \begin{math} K_{particle}=(1/2)M |{\bf v}|^2 \end{math}.
If the particle moves through the fluid, then the fluid exhibits a
dipolar back flow contribution
\begin{math} K_{fluid}=(1/2)\mu |{\bf v}|^2 \end{math}
to the total kinetic energy.

When the sphere is at distance \begin{math} z<\infty  \end{math} from
the wall, the back flow fluid mass current vector must have a zero
component normal to the boundaries. The fluid kinetic energy thereby
depends on \begin{math} z \end{math}. In the limit in which
\begin{math} z>>a \end{math}, the mass matrix
\begin{math} ||M_{jk}(z)|| \end{math}
of the sphere is well known\cite{Milne:1968}; It is
\begin{equation}
\left(
\begin{array}{ccc}
M+\mu_\perp (z) & 0 & 0  \\
0 & M+\mu_\perp(z) & 0  \\
0 &  0 & M+\mu_{||}(z) \\
\end{array}
\right),
\label{fluid4}
\end{equation}
wherein
\begin{eqnarray}
\mu_\perp (z) &=& \mu \left(1+\frac{3}{16}\left(\frac{a}{z}\right)^3\
+\ldots \right),
\nonumber \\
\mu_{||}(z)&=& \mu \left(1+\frac{3}{8}\left(\frac{a}{z}\right)^3
+\ldots \right).
\label{fluid5}
\end{eqnarray}
The potential energy of the sphere induced by fluid momentum fluctuations
is thereby
\begin{equation}
U_{fluid}(z)=\frac{k_BT}{2} \ln \det\left(M^{-1}(z)M^{(0)}\right),
\label{fluid6}
\end{equation}
which reads
\begin{eqnarray}
U_{fluid}(z)&=&-\frac{3\mu k_BT}{8(M+\mu )}\left(\frac{a}{z}\right)^3
+\ldots \ \
\nonumber \\
{\rm with} &\ & \mu = \frac{2\pi \rho a^3}{3} \ \ \
{\rm and}\ \ \ a<<z.
\label{fluid7}
\end{eqnarray}
From fluid velocity thermodynamic fluctuations, it follows that the
sphere will be attracted to wall with a potential proportional to the
temperature and inversely proportional to the third power of the
distance from the wall.

\section{Electric Dipole Moment Fluctuations \label{DipMom}}

If one places a neutral conducting sphere in the neighborhood of a perfectly
conducting wall, then charge rearrangements within the sphere will create
fluctuating electric dipole moments. The fluctuating dipole moments will
induce an attraction between the sphere and the wall as will now be shown.
We again assume that the sphere radius is much less than the distance
between the sphere and the wall \begin{math} a<<z \end{math}. The dipole
moment \begin{math} {\bf d} \end{math} of the conducting sphere will
induce a surface charge on the perfectly conducting wall usually
described in terms of an ``image'' dipole moment
\begin{math} {\bf d}_i \end{math}. The interaction between the dipole moment
and the image is given by
\begin{equation}
\Delta F=
\frac{{\bf d}\cdot {\bf d}_i -3( {\bf d}\cdot{\bf n})({\bf d}_i\cdot{\bf n})}
{16 z^3}.
\label{DipMom1}
\end{equation}
The image dipole moment is related to the dipole moment of the sphere via
\begin{equation}
d_z=d_{iz} ,\ \ d_x=-d_{ix}\ \ {\rm and} \ \ d_y=-d_{iy},
\label{DipMom2}
\end{equation}
so that the interaction free energy reads
\begin{equation}
\Delta F=-\left(\frac{d_x^2+d_y^2+2d_z^2}{16 z^3}\right).
\label{DipMom3}
\end{equation}
Taking the thermodynamic average
\begin{math} U_{dipole}=\overline{\Delta F}  \end{math} yields
\begin{eqnarray}
U_{dipole}(z)&=&-\left(\frac{C_T}{16 z^3}\right)+\ldots
\ \ \ (a<<z)
\nonumber \\
C_T &=& \overline{d_x^2+d_y^2+2d_z^2}
\label{DipMom4}
\end{eqnarray}
wherein the statistical thermodynamic Hamaker constant\cite{Hamaker:1937}
\begin{math} C_T  \end{math} is determined by the polarizability
\begin{math} \alpha_T  \end{math} via
\begin{eqnarray}
k_BT\alpha_T &=& \overline{d_x^2}=\overline{d_y^2}=\overline{d_y^2}
\ \ \ \ \ (a<<z),
\nonumber \\
C_T &=& 4\alpha_T=4a^3\ \ {\rm (conducting\ sphere)}.
\label{DipMom5}
\end{eqnarray}
Altogether, the final attractive potential energy is
\begin{equation}
U_{dipole}(z)=-\frac{k_BT}{4}\left(\frac{a}{ z}\right)^3+\ldots
\ \ \ (a<<z).
\label{DipMom6}
\end{equation}

Note the similarity between the fluid fluctuation potential in
Eq.(\ref{fluid7}) and the dipole fluctuation potential in
Eq.(\ref{DipMom6}). Both potentials obey
\begin{math} U\propto -[k_BT(a/z)^3]  \end{math}
with proportionality constants of similar order unity.
It would at this stage appear that the strength of electrical
and fluid mechanical fluctuation forces are comparable in magnitude. However,
this has only been proven at the classical statistical thermodynamic level of
computation. Let us now consider quantum mechanical fluctuations.

\section{Quantum Fluctuations \label{quant}}

In the quantum mechanical theory of fluctuations, the {\em static} response
function \begin{math} k_BT G_{ij}(Q) \end{math}
at zero temperature in Eq.(\ref{StatThermo2}) is replaced by a complex
frequency \begin{math} \zeta \end{math}
dependent response function \begin{math} k_BT G_{ij}(Q,\zeta )\end{math}
which obeys a dispersion relation with \begin{math} {\Im }m \zeta >0  \end{math}
of the form
\begin{equation}
G_{ij}(Q,\zeta )=
\frac{2}{\pi }\int_0^\infty
\frac{\omega {\Im }m G_{ij}(Q,\omega+i0^+ )d\omega }{(\omega ^2-\zeta^2 )}\ .
\label{quant1}
\end{equation}
The static response function is then the zero frequency limit
\begin{eqnarray}
G_{ij}(Q) &\equiv & \lim_{\zeta \to 0}G_{ij}(Q,\zeta )
\nonumber \\
G_{ij}(Q) &\equiv &
\frac{2}{\pi }\int_0^\infty {\Im }m G_{ij}(Q,\omega+i0^+ )
\frac{ d\omega}{\omega }\ .
\label{quant2}
\end{eqnarray}
The power spectrum of quantum noise corresponding to the frequency dependent
response function \begin{math} G_{ij}(Q,\zeta )\end{math} obeys the
quantum mechanical fluctuation dissipation theorem
\begin{eqnarray}
&&\frac{1}{2}\left<\{\Delta X_i(t),\Delta X_j(0)\}\right>
=\int_{-\infty}^\infty S_{ij}(Q,\omega )e^{-i\omega t}d\omega ,
\nonumber \\
&&S_{ij}(Q,\omega )=\frac{E_T(\omega )}{\pi \omega }
{\Im }m G_{ij}(Q,\omega+i0^+ ),
\nonumber \\
&&E_T(\omega )=\left(\frac{\hbar \omega }{2 }\right)
\coth\left(\frac{\hbar \omega}{2k_BT}\right).
\label{quant3}
\end{eqnarray}
Employing an identity,
\begin{eqnarray}
\frac{E_T(\omega )}{k_BT} &=&
\sum_{n=-\infty}^\infty \frac{\omega^2}{\omega^2+\omega_n^2},
\nonumber \\
\omega_n &=& \frac{2\pi k_BTn}{\hbar}\ ,
\label{quant4}
\end{eqnarray}
along with Eqs.(\ref{quant1}), (\ref{quant3}) and (\ref{quant4}) imply
\begin{eqnarray}
&\ & \frac{1}{2}\left<\Delta X_i\Delta X_j+\Delta X_j\Delta X_i\right>
\nonumber \\
&\ &=k_BT\sum_{n=-\infty}^\infty G_{ij}(Q,i|\omega_n| ).
\label{quant5}
\end{eqnarray}
It is worthwhile to compare the quantum mechanical fluctuation Eq.(\ref{quant5}) to
the classical Eq.(\ref{StatThermo2});
\begin{equation}
\overline{\Delta X_i \Delta X_j}=k_BT G_{ij}(Q,0)\equiv k_BT G_{ij}(Q).
\label{quant6}
\end{equation}

For a single fluctuating variable,
say \begin{math} X=\sum_i a_iX_i  \end{math},
one obtains from Eq.(\ref{quant5}) the expression
\begin{eqnarray}
\left<\Delta X^2\right> &=&
k_BT\sum_{n=-\infty}^\infty G(Q,i|\omega_n| ),
\nonumber \\
G(Q,\zeta )&=&\frac{2}{\pi }
\int_0^\infty \frac{ \omega {\Im }m G_{ij}(Q,\omega+i0^+ )d\omega}
{\omega^2-\zeta^2 }\ .
\label{quant7}
\end{eqnarray}
Employing an inequality,
\begin{equation}
\frac{1}{\omega^2+\omega_n^2}\le \frac{1}{\omega_n^2},
\label{quant8}
\end{equation}
and a definition for the \begin{math} X-  \end{math}motion
frequency scale \begin{math} \omega_\infty  \end{math},
\begin{equation}
\omega_\infty ^2=\left(\frac{2}{\pi G(Q,0)}\right)
\int_0^\infty\omega{\Im}mG(Q,\omega+i0^+)d\omega ,
\label{quant9}
\end{equation}
into Eq.(\ref{quant7}) yields the upper bound
\begin{eqnarray}
\left|\left<\Delta X^2\right>-k_BTG(Q,0)\right|
\nonumber \\
\le 2\omega_\infty ^2 k_BTG(Q,0)\sum_{n=1}^\infty \frac{1}{\omega_n^2}\ ,
\label{quant10}
\end{eqnarray}
i.e.
\begin{equation}
\left|\frac{\left<\Delta X^2\right>}{k_BTG(Q,0)}-1\right|
\le \frac{1}{12}\left(\frac{\hbar \omega_\infty }{k_BT}\right)^2 .
\label{quant11}
\end{equation}
From the inequality in Eq.(\ref{quant11}) we find that a sufficient
condition for employing {\em classical} fluctuations is
\begin{equation}
\left<\Delta X^2\right>\approx k_BT G(Q,0)
\ \ {\rm if}\ \ \hbar \omega_\infty <<k_BT
\label{quant12}
\end{equation}
in agreement the classical Eq.(\ref{StatThermo2}).

\section{Fluctuation Frequencies \label{frequency}}

Our purpose is to estimate \begin{math} \omega_\infty \end{math}
for both the fluid mechanical and the electrical fluctuation forces.
We conclude for ``room temperature'' that
\begin{eqnarray}
\hbar \omega_\infty &<<& k_BT\ \ \ {\rm fluid\ mechanics\ (classical)},
\nonumber \\
\hbar \omega_\infty &>>& k_BT\ \ \ {\rm dipole\ moments\ (quantum)}.
\label{frequency_result}
\end{eqnarray}
The derivations follow.

\subsection{Effective Mass Sum Rule}

If \begin{math} {\bf p} \end{math} denotes the momentum of a
colloidal particle within a fluid, then the dynamical mass of the
particle obeys the Kubo formula
\begin{equation}
{\sf M}_{ij}(\zeta )=\frac{i}{\hbar}\int_0^\infty
\left<\left[p_i(t),p_j(0)\right]\right>e^{i\zeta t}dt.
\label{frequency1}
\end{equation}
From Eq.(\ref{frequency1}) one finds
\begin{eqnarray}
&\ &\frac{i}{\hbar}\left<\left[p_i(t),p_j(0)\right]\right>
\nonumber \\
&=&\frac{2}{\pi }\int_0^\infty{\Im m}{\sf M}_{ij}(\omega +i0^+ )
\sin(\omega t)d\omega ,
\label{frequency2}
\end{eqnarray}
from which follows the equal time commutation sum rule
\begin{equation}
\frac{i}{\hbar}\left<\left[\dot{p}_i,p_j\right]\right>=\frac{2}{\pi }
\int_0^\infty \omega {\Im m}{\sf M}_{ij}(\omega +i0^+ ) d\omega .
\label{frequency3}
\end{equation}
If the microscopic force \begin{math} \dot{\bf p}={\bf f} \end{math} on
the colloidal particle is derivable from a potential
\begin{math} {\bf f}=-{\bf grad}V \end{math}, then the sum rule
Eq.{\ref{frequency3}} obeys
\begin{equation}
\frac{2}{\pi }\int_0^\infty \omega {\Im m}{\sf M}(\omega +i0^+ )d\omega
=\left<{\bf grad\ grad}V\right>.
\label{frequency4}
\end{equation}
On the other hand, from the dispersion relation,
\begin{equation}
{\sf M}(\zeta )=
\frac{2}{\pi }\int_0^\infty
\frac{\omega {\Im }m {\sf M}(\omega+i0^+ )d\omega }{(\omega ^2-\zeta^2 )}
\ \ \ {\rm if}\ \ \ {\Im }m \zeta >0,
\label{frequency5}
\end{equation}
it follows that the static mass obeys
\begin{equation}
{\sf M}=
\frac{2}{\pi }\int_0^\infty
{\Im m} {\sf M}(\omega+i0^+ )\frac{d\omega }{\omega }.
\label{frequency6}
\end{equation}
From Eqs.(\ref{frequency4}) and (\ref{frequency6}) one computes the
Hooks law frequency tensor
\begin{equation}
{\bf \Omega}_\infty ^2={\sf M}^{-1}\cdot \left<{\bf grad\ grad}V\right>.
\label{frequency7}
\end{equation}
For a given principle ``X-direction'' of the tensor, the hydrodynamic
frequency scale is given by
\begin{equation}
{\bf \omega}_\infty ^2=\frac{1}{M}
\left<\frac{\partial^2 V}{\partial X^2}\right>.
\label{frequency8a}
\end{equation}

The mass of the colloidal particle is proportional to the volume of the
particle. The interaction \begin{math} V \end{math} between the colloidal
particle and the fluid is spatially non-zero only in the neighborhood of
the particle surface. Thus,
\begin{math} \left<\partial ^2 V/\partial X^2\right> \end{math}
is proportional to the contact surface area. If the number of atoms
within the colloidal particle is denoted by \begin{math} N \end{math},
then the number of atoms on the colloidal particles surface is proportional
to \begin{math} N^{2/3} \end{math}. The frequency in Eq.(\ref{frequency8a})
may be estimated by
\begin{equation}
{\bf \omega}_\infty ^2\sim N^{-1/3}\omega_{vib}^2
\label{frequency8b}
\end{equation}
where \begin{math} \omega_{vib} \end{math} is a typical
atomic vibrational (say phonon) frequency. As a numerical example,
let us consider a colloidal particle with
\begin{math} N\sim 10^{10} \end{math} and with a vibrational frequency
obeying \begin{math} \hbar \omega_{vib}\sim k_BT \end{math}
at room temperature. For such a colloidal particle
\begin{math} \hbar \omega_\infty \sim 0.02 \ k_BT \end{math}
which is in the classical regime of Eq.(\ref{frequency_result}).

\subsection{Polarizability Sum Rule}

If \begin{math} {\bf d} \end{math} denotes the electric dipole moment
of a colloidal particle within a fluid, then the polarizability of the
particle obeys
\begin{equation}
\alpha_{ij}(\zeta )=\frac{i}{\hbar}\int_0^\infty
\left<\left[d_i(t),d_j(0)\right]\right>e^{i\zeta t}dt.
\label{frequency9}
\end{equation}
From Eq.(\ref{frequency9}) one finds
\begin{eqnarray}
&\ &\frac{i}{\hbar}\left<\left[d_i(t),d_j(0)\right]\right>
\nonumber \\
&=& \frac{2}{\pi }\int_0^\infty{\Im m}\alpha_{ij}(\omega +i0^+ )
\sin(\omega t)d\omega ,
\label{frequency10}
\end{eqnarray}
from which follows the equal time commutation sum rule
\begin{equation}
\frac{i}{\hbar}\left<\left[\dot{d}_i,d_j\right]\right>=\frac{2}{\pi }
\int_0^\infty \omega {\Im m}\alpha_{ij}(\omega +i0^+ ) d\omega .
\label{frequency11}
\end{equation}
The electric dipole moment and its rate of change,
summed over all the charges within the colloidal particle,
is given by
\begin{eqnarray}
{\bf d} &=& e\sum_k z_k {\bf r}_k,
\nonumber \\
\dot{\bf d} &=& e\sum_k z_k \dot{\bf r}_k
=e\sum_k z_k {\bf v}_k.
\label{frequency12}
\end{eqnarray}
From the equal time commutation relation
\begin{math}
[{\bf v}_k,{\bf r}_l]=-i\hbar \delta_{kl}({\sf 1}/M_k)
\end{math}
one finds that
\begin{math}
[\dot{d}_i,d_j]=-i\hbar \delta_{ij}\sum_k(e^2z_k^2/M_k)
\end{math}
yielding
\begin{equation}
\frac{2}{\pi }\int_0^\infty \omega
{\Im m} \alpha_{ij}(\omega+i0^+ )d\omega =
\delta_{ij}\sum_k\left(\frac{e^2z_k^2}{M_k}\right).
\label{frequency13}
\end{equation}
From the dispersion relation (with \begin{math} {\Im }m \zeta >0 \end{math})
\begin{equation}
\alpha_{ij}(\zeta )=
\frac{2}{\pi }\int_0^\infty
\frac{\omega {\Im }m \alpha_{ij}(\omega+i0^+ )d\omega }{(\omega ^2-\zeta^2 )},
\label{frequency14}
\end{equation}
it follows that the static polarizability obeys
\begin{equation}
\alpha_{ij}\equiv \alpha_{ij}(0)=
\frac{2}{\pi }\int_0^\infty
{\Im m} \alpha_{ij}(\omega+i0^+ )\frac{d\omega }{\omega }.
\label{frequency15}
\end{equation}
For a spherical colloidal particle,
\begin{equation}
\omega_\infty ^2=\left(\frac{2}{\pi \alpha_T}\right)
\int_0^\infty\omega{\Im}m \alpha (\omega+i0^+)d\omega.
\label{frequency16}
\end{equation}
For a conducting sphere of volume
\begin{math} V=(4\pi a^3/3) \end{math} the frequency scale
\begin{equation}
\omega_\infty ^2=
\frac{e^2}{\alpha_T }\sum_k\left(\frac{z_k^2}{M_k}\right)
=\frac{4\pi e^2}{3V}\sum_k\left(\frac{z_k^2}{M_k}\right).
\label{frequency17a}
\end{equation}
The plasma frequency for a portion of condensed matter
\begin{equation}
\Omega_p^2
=\frac{4\pi e^2}{V}\sum_k\left(\frac{z_k^2}{M_k}\right),
\label{frequency17b}
\end{equation}
is dominated by electronic oscillations
\begin{math} \Omega_p^2\approx (4\pi ne^2/m) \end{math}
wherein \begin{math} m \end{math} and \begin{math} n \end{math}
represent, respectively, the electron mass and density of
electrons per unit volume. The frequency scale for dipolar
fluctuations is then
\begin{equation}
\omega_\infty ^2 =\frac{\Omega_p^2}{3}\ .
\label{frequency18}
\end{equation}
A metallic plasma frequency is of order
\begin{math} (\hbar \Omega_p/k_B)\sim 10^5 \ ^\circ K  \end{math}.
Eq.(\ref{frequency18}) then implies
\begin{math} \hbar \omega_\infty >> k_BT  \end{math},
as in Eq.(\ref{frequency_result}), for temperatures near
room temperature.

\section{Quantum Dipolar Forces \label{QDF}}

It has been found at room temperature that fluid fluctuation forces are
classical and electric fluctuation forces are quantum mechanical. An estimate
of the quantum mechanical dipolar potential follows. The energy of
interaction between a conducting sphere and wall due to dipole quantum
fluctuations is found by summing the polarizability over
Matsubara frequencies\cite{Matsubara:1955} as in Sec.\ref{quant}; i.e.
\begin{equation}
U_{dipole}(z)=-\left (\frac{k_BT}{4 z^3}\right)
\sum_{-\infty}^\infty \alpha(i|\omega_n|).
\label{QDF1}
\end{equation}

A simple model for the polarizability will be found in order to estimate
the potential energy in Eq.(\ref{QDF1}). We assume that the polarizability
has a single pole at frequency \begin{math} \omega_\infty  \end{math}; i.e.
\begin{equation}
\alpha(\zeta)\approx {\alpha_T\omega_\infty^2 \over \omega_\infty^2 - \zeta^2}.
\label{QDF2}
\end{equation}
The residue at the pole has been fixed so that
\begin{math} \alpha (0)\equiv \alpha_T \end{math}.
Substituting Eq.(\ref{QDF2}) in Eq.(\ref{QDF1}) one finds for the
interaction potential
\begin{eqnarray}
U_{dipole}(z)&\approx &-\left(\frac{k_BT}{4 z^3}\right)\sum_{n=-\infty}^\infty
{\alpha_T\omega_\infty^2 \over \omega_\infty^2 +\omega_n^2}
\nonumber \\
U_{dipole}(z)&\approx &
-\left(\frac{\hbar\omega_\infty\alpha_T}{8 z^3}\right)
\coth(\frac{\hbar\omega_\infty}{2k_BT}).
\label{QDF3}
\end{eqnarray}
Since \begin{math} \hbar\omega_\infty \gg k_BT \end{math} one finds
\begin{equation}
U_{dipole}(z)=-\frac{\hbar\omega_\infty\alpha_T}{8 }
\left(\frac{1}{z}\right)^3
=-\frac{\hbar\omega_\infty }{8 }\left(\frac{a}{z}\right)^3.
\label{QDF4}
\end{equation}

By comparing Eq.(\ref{QDF4}) to Eq.(\ref{fluid7}), it seems that the
dipole fluctuation forces are much larger than the fluid fluctuation
forces we find that
\begin{equation}
\frac{U_{fluid}(z)}{U_{dipole}(z)}=
\left[\frac{3\mu }{M+\mu }\right]\frac{k_BT}{\hbar \omega_\infty}
\label{QDF5}
\end{equation}
The term on the right hand side of Eq.(\ref{QDF5}) which is in square
brackets is of order unity. Thus
\begin{equation}
\frac{U_{fluid}(z)}{U_{dipole}(z)}\sim
\frac{k_BT}{\hbar \omega_\infty} <<1.
\label{QDF6}
\end{equation}
Using typical plasma frequencies for metals,
\begin{math} \omega_\infty = (\Omega_p/\sqrt{3}) \sim (10^{16}/sec) \end{math}
or equivalently
\begin{math} (\hbar \omega_\infty/k_B) \sim 10^5 \ ^\circ K   \end{math}.
The inequality in Eq.(\ref{QDF6}) holds by a very large margin.

\section{Fluid Motion \label{fm}}

The static classical fluid force between two spherical colloidal particles
separated by \begin{math} r \end{math} may be shown to be derived
from the potential
\begin{equation}
U_{fluid}(r)=-3\pi ^2\left[\frac{k_BT\rho^2a^6}{(\mu+M)^2}\right]
\left(\frac{a}{r}\right)^6\ \ (r>>a),
\label{fm1}
\end{equation}
wherein \begin{math} \rho \end{math} is the fluid mass density and
\begin{math} \mu \end{math} is the Euler mass and \begin{math} M \end{math}
is the bare mass of the colloidal particle.
The above potential is derived from a long time scale statistical averaging
over those fluid mechanical fluctuations which otherwise induce colloidal
particle Brownian motion. Direct observation\cite{Meiners:1999,Henderson:2002}
of Brownian motion forces require shorter time scales. Typical
experimental bandwidths for micron scale colloidal particle size are about
a tenth of a megahertz.

For two identical metallic colloidal particles separated by \begin{math} r \end{math},
the quantum electric field fluctuation potential is given by
\begin{eqnarray}
&\ &U_{dipole}(r)=-\frac{3k_BT}{r^6}
\sum_{n=-\infty}^{\infty}\alpha(i|\omega_n|)\alpha(i|\omega_n|)
\nonumber \\
&\ &{\rm for} \ \ r>>a.
\label{fm2}
\end{eqnarray}
In the single pole approximation for
\begin{math} \alpha (\zeta ) \end{math}
with the metallic \begin{math} \alpha (0)=a^3 \end{math}
we find that
\begin{eqnarray}
&\ &U_{dipole}(r)\approx
-\left[\frac{3\hbar \omega_\infty}{2}\right]\left(\frac{a}{r}\right)^6
\nonumber \\
&\ &{\rm for}\ \ r>>a \ \ {\rm and}\ \ \hbar \omega_\infty >>k_BT.
\label{fm3}
\end{eqnarray}
Note that
\begin{equation}
\frac{U_{fluid}(r)}{U_{dipole}(r)}\sim
\left[\frac{\rho^2a^6}{(\mu+M)^2}\right]
\left[\frac{k_BT}{\hbar \omega_\infty}\right]<<1.
\label{fm4}
\end{equation}
The very large magnitude of
\begin{math} \omega_\infty \sim (10^{16} /sec)  \end{math}
forbids direct observation of dipole fluctuations.

\section{Conclusions \label{conc}}

The general theory of quantum mechanical fluctuations
was discussed with particular emphasis on computing the
frequency scales of motion from sum rules. The frequency scales
determine whether the fluctuations and thereby the forces are
classical or quantum mechanical in nature.

We have shown how fluctuations in fluid velocity fields and in
electric fields give rise to forces exerted on colloidal particles.
Fluctuation forces were computed in detail for the case of colloidal
particles attracted to the walls of the suspension container. The resulting
van der Waals force has the form
\begin{equation}
U_{Wall}(z)=-U_1\left(a/z\right)^3\ \ {\rm for}\ \ z>>a.
\label{conc1}
\end{equation}
The electric field fluctuation contribution
to \begin{math} U_1 \end{math}  dominates the fluid mechanical
contribution to \begin{math} U_1 \end{math} as in Eq.(\ref{QDF6}).
The long ranged static attraction between two spherical
particles of radius \begin{math} a \end{math} separated by
a distance \begin{math} r \end{math} has the van der Waals form
\begin{equation}
U_{particle}(r)=-U_0\left(a/r\right)^6\ \ {\rm for}\ \ r>>a.
\label{conc1}
\end{equation}
The electric field fluctuation contribution
to \begin{math} U_0 \end{math}  dominates the fluid mechanical
contribution to \begin{math} U_0 \end{math} as in Eq.(\ref{fm4}).

At the level of classical statistical thermodynamics,
the fluid velocity and electric field contributions to
this potential are comparable. When quantum fluctuation effects
are taken into account, the electric fluctuation contribution
to the potentials dominates the fluid mechanical
contribution to the potentials as in Eqs.(\ref{QDF6})
and (\ref{fm4}). The electric field fluctuation long ranged
forces can be, and have been\cite{Tao:1989,Woestman:1993}, observed
by measuring phase separations in some colloidal suspensions.
Further work on colloidal forces in still other geometries would
be of general interest.

\end{document}